\documentclass[twocolumn,showpacs,preprintnumbers,amsmath,amssymb,pra]{revtex4}
\usepackage{epsfig}                                                            
\usepackage{graphicx}
\usepackage{dcolumn}
\usepackage{color}
\usepackage{bm}
\begin{document}

\title {\bf Quantum time of flight distribution for cold trapped atoms}

\author{Md. Manirul Ali\footnote{mani@bose.res.in}$^1$,
 Dipankar Home\footnote{dhome@bosemain.boseinst.ac.in}$^2$,
A. S. Majumdar\footnote{archan@bose.res.in}$^1$,
and Alok K. Pan\footnote{apan@bosemain.boseinst.ac.in}$^2$}

\affiliation{$^1$ S. N. Bose National Centre for Basic Sciences, Salt Lake, 
Calcutta 700098, India}

\affiliation{$^2$Department of Physics, Bose Institute, Calcutta
700009, India}

\date{\today}

\begin{abstract}
The time of flight distribution for a cloud of cold atoms falling 
freely under gravity is considered. We generalise the probability current 
density approach to calculate the quantum arrival time distribution for the 
mixed state describing the Maxwell-Boltzmann distribution of velocities for 
the falling atoms. We find an empirically testable difference between the 
time of flight distribution calculated using the quantum probability current 
and that obtained from a purely classical treatment which is usually 
employed in analysing time of flight measurements. The classical time of 
flight distribution matches with the quantum distribution in the large mass 
and high temperature limits.  
\end{abstract}        

\maketitle                                                                    
             
\section{Introduction}

In recent times laser cooling and trapping of atoms has become an 
area of active research\cite{expt}. The measurement of the initial 
temperature of the 
cloud of atoms is crucial for characterising the properties of atom traps.
The temperature of the cloud can be inferred from the velocity distribution
of atoms in the cloud. A well-known technique of measuring this velocity 
distribution is the time of flight (TOF) method. Measurements of the TOF 
distribution have 
been employed to analyse various experimental data such as those involving
ions and isotopes\cite{ion}, and also in performing mass spectroscopy of
biomolecules like DNA\cite{dna}. 

The theoretical treatment
of the TOF distribution that can be obtained using for instance, the 
Green's function 
method\cite{weiss},
however, turns out to produce perfect agreement with the TOF distribution
obtained by using Newton's equations for ballistic motion of particles 
accelerated 
by the earth's gravitational field\cite{yavin}.  
Thus, the interpretation of the results of the various TOF 
experiments\cite{expt,ion,dna} where classical trajectories are 
inferred from  Newtonian mechanics\cite{tof} remains debatable,
especially in the domain of small atomic masses and low temperatures where
quantum mechanical effects should be significant.

Though there exists no unique prescription for the definition of 
time of flight and arrival time in quantum 
mechanics,  experimentalists measure arrival times
of elementary particles, atoms and molecules using the TOF methods. 
In spite of the 
difficulties to give time an
observable status in quantum mechanics, several logically consistent schemes 
for the
treatment of the arrival time 
distribution
have been formulated, such as those based on axiomatic 
appraches\cite{kijowski},
opearator constructions\cite{operator}, and trajectory
models\cite{bohmarrtim}. It is thus desirable 
that some of the
conceptually sound theoretical formulations of the quantum mechanical arrival 
time distribution\cite{1} be confronted with accurate experimental data.
If such quantum mechanical approaches are employed for analysing experiments 
using TOF measurements,
it should not only enable to determine the empirical viability of various
competing
arrival time models\cite{1}, but also possibly shed new light on the
conventional interpretation of the results of these experiments. 

In this paper we employ the probability current approach\cite{current} 
towards obtaining
the quantum arrival time distribution of cold trapped atoms.
The 
probability current approach for computation of the mean arrival time of a 
quantum
ensemble not only provides an unambiguous definition of arrival time at the 
quantum
mechanical level\cite{current,holland,ali}, but also adresses the issue of 
obtaining
the proper classical limit of the time of flight of massive quantum particles
\cite{class,weq}. Here we derive the quantum arrival time distribution
for the case of initially trapped atomic clouds that are subsequently 
allowed to
fall freely under gravity\cite{weiss,lett}. We compute the mean time of 
arrival for these atoms and compare it with the mean arrival time
obtained through the classical time of flight analysis\cite{yavin} that
has frequently been employed for such experiments\cite{expt,ion,dna}.
Our analysis predicts the mass and temperature range of the atomic clouds
where the quantum mechanical treatment alters the arrival time distribution
and the mean arrival time from that obtained through the classical analysis.

\section{Classical analysis of time of flight measurements}

We begin with a brief description of the classical analysis of TOF 
measurements of trapped atoms. A probe 
laser, focussed in the form of a sheet, is placed underneath the atomic cloud.
When the trapping forces are turned off, the cold atom cloud falls through
the laser probe under the influence of gravity. It is then possible to
detect the fluorescence from the atoms as they reach the sheet. The
fluorescence is measured as a function of time and the initial temperature
of the cloud is determined by fitting the experimental result to the 
theoretically predicted TOF signal of the cloud\cite{weiss,lett}.
A detailed derivation of the TOF signal recorded by the detector (that is,
the number of atoms arriving at the probe laser as a function of time) was
derived by Yavin {\it et al}\cite{yavin}. 

The cloud of atoms
consisting of noninteracting particles has a Maxwell-Boltzmann 
velocity distribution given (in one dimension) by 
\begin{equation}
\Pi(v) dv= {\left(\frac{m}{2 \pi k T} \right)}^{1/2} 
\exp\left(-\frac{m v^2}{2 k T}\right) dv
\end{equation}
where $T$ is the initial temperature of the cloud, and $m$ is the atomic mass. 
Using the Newton's equations for ballistic motion of a particle accelerated 
by the earth's gravitational field (in the vertical $z$-direction), the 
velocity is obtained in terms of the time of flight as
\begin{equation}
v=(z+\frac{1}{2}g t^2)/t
\end{equation}
Substituting the above expression for $v$ from Eq.(2) in Eq.(1), one can
obtain the time of flight distribution at an arbitrary distance $z$, given by 
\begin{eqnarray}
\nonumber
\Pi_{C}(t)dt &=& {\left(\frac{m}{2 \pi k T} \right)}^{1/2}
\exp\left(-\frac{m(z+\frac{1}{2}g t^2)^2}{2 k T t^2} \right)\\
&&\times \frac{(-z+\frac{1}{2}g t^2)}{t^2}dt
\label{classdist}
\end{eqnarray}
The corresponding classical mean time of flight or mean arrival
time $\overline{\tau_C}$ for the atomic cloud calculated using 
${\Pi_{C}}(t)$
as the time of flight distribution is given by
\begin{eqnarray}
\overline{\tau_C}=\frac{\int_{0}^{\infty}
{\Pi_{C}}(t) t\, dt}{\int _{0}^{\infty}
{\Pi_{C}}(t) dt}
\label{classarrtim}
\end{eqnarray}
For simplicity, here we restrict ourselves to the case of a point-sized cloud. 
The three-dimensional calculation was done by Yavin et al\cite{yavin} using 
a simple coordinate
transformation and the same expression was obtained for the TOF distribution. 
These authors\cite{yavin} have claimed perfect agreement of their results
with a previous calculation\cite{weiss} where
the TOF distribution was derived using a sophisticated Green's function 
technique. 
We call this TOF distribution given by Eq.(\ref{classdist}) as the classical 
time of flight distribution, and Eq.(\ref{classarrtim}) denotes the
corresponding classical mean arrival time
since the classical Newtonian equation is used from the outset to derive 
this TOF distribution. 

\section{The quantum arrival time distribution through the probability current}

Our aim here is to derive an expression for the time of flight distribution for
the atomic cloud through the quantum probability current without using any 
classical ingredients. To that end, let the initial 
state of each of the atoms be represented by a one dimensional 
Gaussian wave function of the form
\begin{eqnarray}
\psi(z,0)=(2\pi {\sigma_0}^2)^{-1/4} \exp(\frac{imv}{\hbar}z) \exp\left(-\frac{z^2}
{4{\sigma_0}^2}\right)
\label{wavefn1}
\end{eqnarray}
centered at $z=0$ and moving
with a group velocity $v$. The Schr${\ddot o}$dinger time evolved wave 
function 
under the Hamiltonian $H=p^2/2m +mgz$ is given by
\begin{eqnarray}
\nonumber
&& \psi(z,t) = \left(2\pi s^{2}_{t}\right)^{-1/4}
\exp \left[ \frac{ \left( z- vt + \frac{1}{2}g t^2\right)^2}{4s_{t}\sigma_{0}} \right]\\
&& \times \exp \left[i(\frac{m}{\hbar}) \left\{\left(v- gt\right) \left(z-vt/2\right)
-\frac{1}{6}g^2 t^3 \right\} \right]
\end{eqnarray}
where $s_{t}=\sigma_{0}\left(1+i\hbar t/2{m}\sigma_{0}^{2}\right)$. 

Considering the free fall of the atoms  
under
gravity, the expression for the Schr${\ddot o}$dinger  probability current 
density
\begin{equation}
J(z,t) \equiv \frac{i\hbar}{2m}(\psi \frac{\partial \psi^{\ast}}{\partial z}
-\psi^{\ast}\frac{\partial \psi}{\partial z})
\label{currdens}
\end{equation}  
for the time evolved state is 
calculated 
using the initial state given by
Eq.(\ref{wavefn1}) to be
\begin{eqnarray}
J(z,t)=P(z,t) \left[(v-gt)+ \frac{{\hbar}^2 t}{4 m^2 {\sigma_0}^2 {\sigma}^2}( z- vt +
\frac{1}{2}g t^2)\right]
\label{purecurrent}
\end{eqnarray}
where the expression for
the position probability distribution is given by
\begin{eqnarray}
P(z,t)=\frac{1}{(2 \pi {\sigma}^2)^{1/2}} \exp \left[-\frac{\left( z- vt + 
\frac{1}{2}g t^2\right)^2}{2 {\sigma}^2} \right]
\label{pure}
\end{eqnarray}
with $\sigma=\sigma_{0}\left(1+\hbar^{2}t^{2}/4{m}^{2}\sigma_{0}^{4}
\right)^{1/2}$. 
The modulus of the probability current density $J(z,t)$ given by 
Eq.(\ref{purecurrent}) provides the arrival time distribution for a pure 
wave packet
falling under gravity.
Note that the quantum probability current as defined by Eq.(\ref{currdens})
is formally ambiguous up to
a total divergence term\cite{finkelstein}. However, $J(z,t)$ can be uniquely 
defined through relativistic
wave equations which impart appropriate spin-dependent corrections
to it that persist
even in the non-relativistic limit\cite{holland,baere}. The ensuing 
arrival time distribution defined through the probability current thus  
contains a spin-dependent correction for particles with spin\cite{ali}.

The atomic cloud is represented by an ensemble of particles in 
thermal equilibrium with a  thermal distribution of initial 
velocities. Each particle has a wave function of the form (\ref{wavefn1}), 
with a Maxwell-Boltzmann distribution of initial velocities given by Eq.(1). 
Thus the initial thermal state of the atomic cloud we have described 
is a mixed state. We obtain the corresponding position probability 
distribution by averaging the pure state distribution (\ref{pure}) 
over a 
thermal distribution 
of initial velocities. The result is
\begin{eqnarray}
\nonumber
P_{T}(z,t) &=& {\left(\frac{m}{2 \pi k T} \right)}^{1/2} \int_{-\infty}^{\infty} P(z,t)
\exp\left(-\frac{m v^2}{2 k T}\right) dv \\
&=& \frac{1}{(2 \pi {\sigma_T}^2)^{1/2}} \exp \left[-\frac{\left(z+\frac{1}{2}g t^2
\right)^2}{2 {\sigma_T}^2} \right] 
\end{eqnarray}  
where ${\sigma_T}^2={\sigma}^2 + ({kT}/m) t^2$. The peak of the position 
probability
distribution $P_{T}(z,t)$ follows the classical trajectory and the effect 
of the mass and
temperature dependences of the position probability occurs essentially because
the spreading of the wave packet is different for different atomic mass and 
temperature 
of the cloud.

\begin{figure}
\centering
\epsfxsize=3.5in\epsfysize=4.5in
\rotatebox{0}{\epsfbox{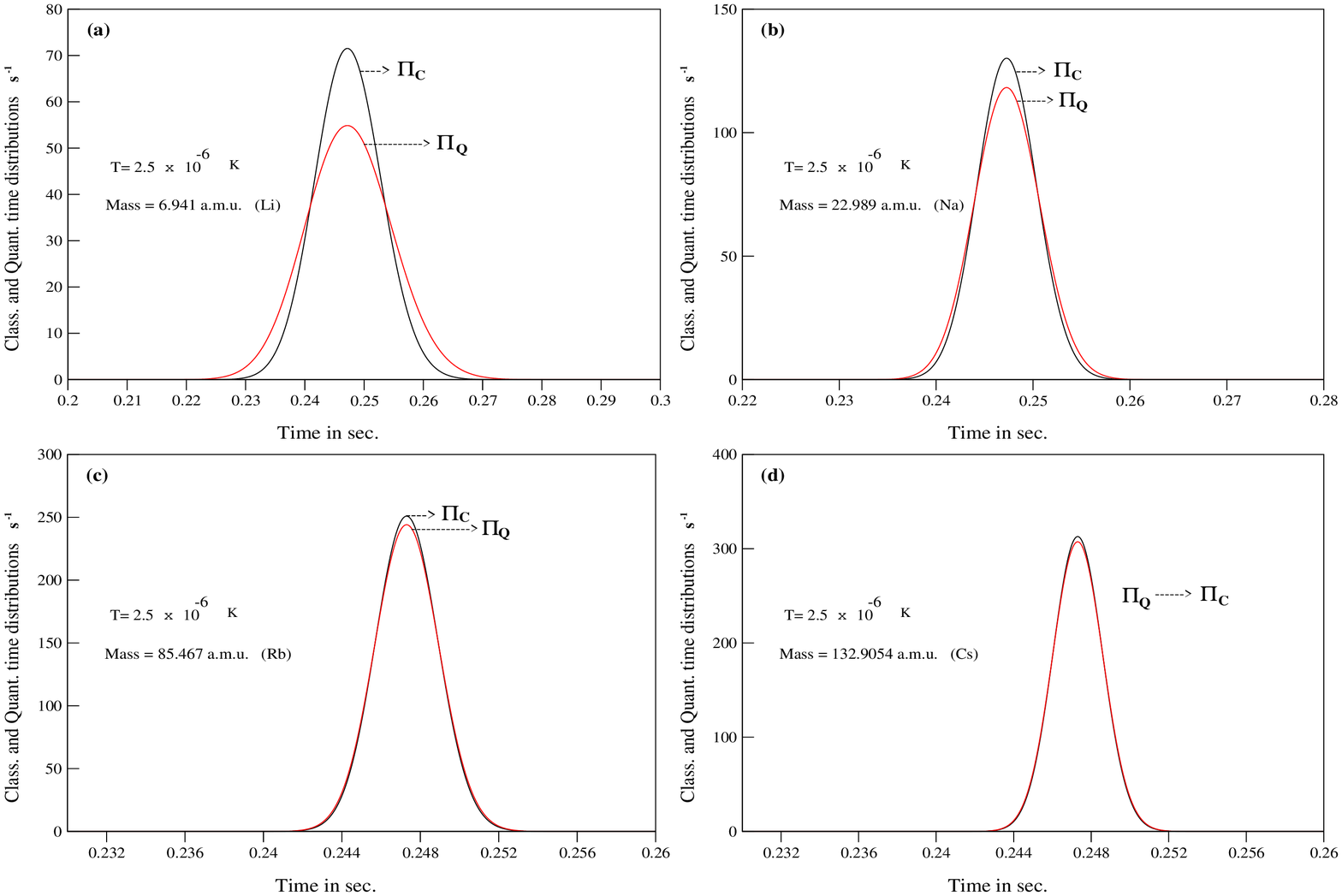}}
\caption{The classical (${\Pi_{C}}(t)$) and quantum (${\Pi_{Q}}(t)$)
TOF distributions of the atomic cloud falling freely under gravity are plotted 
for varying mass of the atoms at a fixed temperature $T=2.5 \times 10^{-6} K$ 
with
$\sigma_0=10^{-5} cm$, $Z=-30 cm$.}
\label{fig.1}
\end{figure}

Taking the modulus of the quantum probability current density as determining
the quantum arrival time distribution\cite{current}, we obtain the 
arrival time probability 
distribution 
for the atomic cloud (which is a mixed state) by averaging the pure state 
distribution (\ref{purecurrent}) 
over a thermal distribution of initial velocities. 
The result is
\begin{eqnarray}
\nonumber
{\Pi_{Q}}(t) &=& \left| {\left(\frac{m}{2 \pi k T} \right)}^{1/2} \int_{-\infty}^{\infty} J(z,t)
\exp\left(-\frac{m v^2}{2 k T}\right) dv \right|\\
\nonumber
&=& \frac{1}{(2 \pi {\sigma_T}^2)^{1/2}} \exp \left[-\frac{\left(z+\frac{1}{2}g t^2
\right)^2}{2 {\sigma_T}^2} \right]\\
&& \times \left[\frac{(z+\frac{1}{2}g t^2)(\frac{kT}{m}+\frac{\hbar^2}{4 m^2 
{\sigma_0}^2})t}{{\sigma_T}^2} \right]
\label{quantdistri}
\end{eqnarray}
In this way we generalise the probability current density approach to 
calculate the 
quantum arrival time distribution for the mixed state at finite temperature. 
It may be mentioned here that in the present calculation we neglect the
small spin-dependent correction\cite{ali} that may 
appear in the probability 
current density, as mentioned above, and consequently
the mean arrival time that we compute for any fermionic atoms.

\begin{figure}
\centering
\epsfxsize=3.5in\epsfysize=4.5in
\rotatebox{0}{\epsfbox{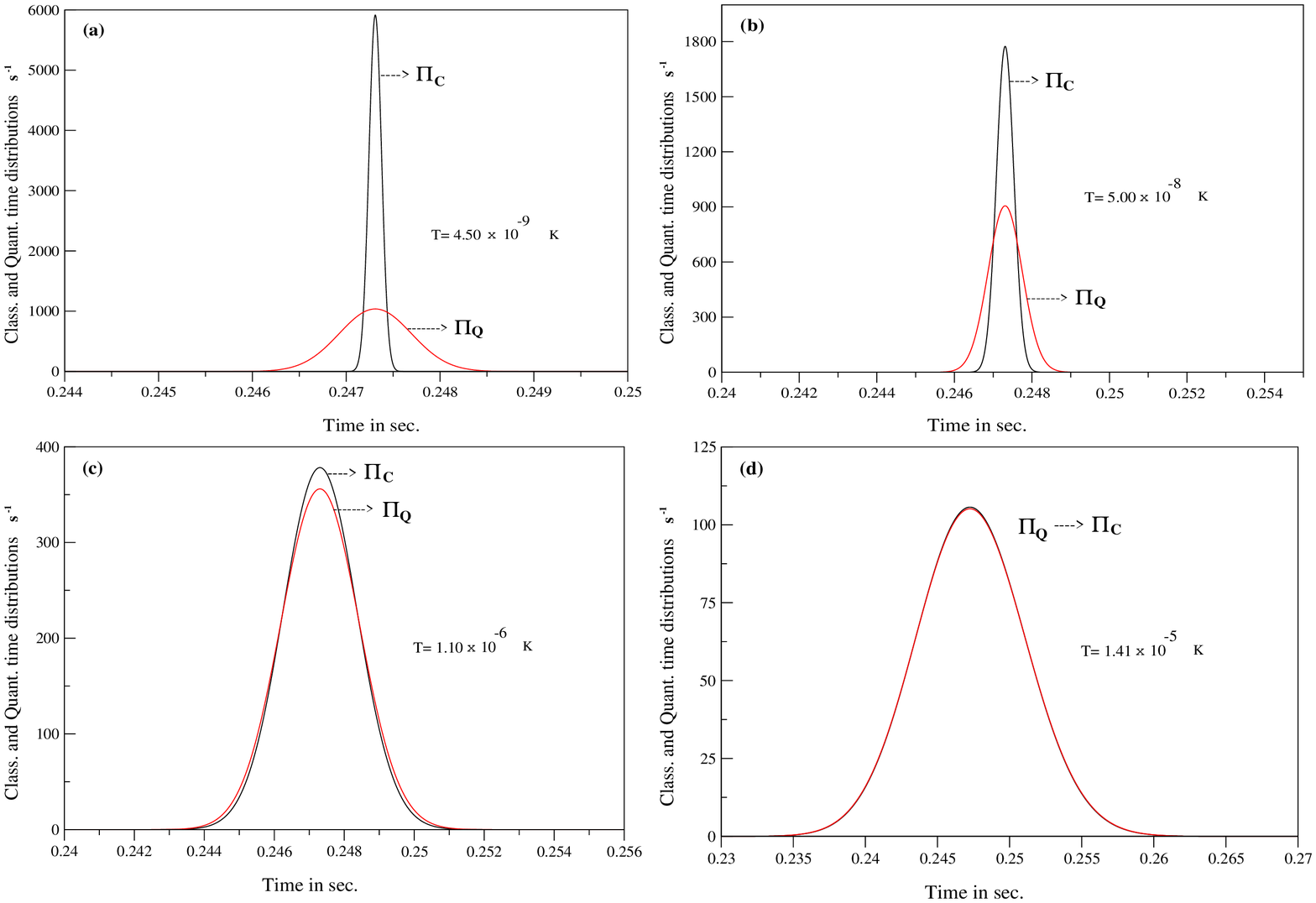}}
\caption{The classical (${\Pi_{C}}(t)$) and quantum (${\Pi_{Q}}(t)$)
TOF distributions of the atomic cloud falling freely under gravity are 
plotted for
varying temperatures at a fixed mass of Rb atom ($m=85.4678 a.m.u$)  
with $\sigma_0=10^{-5} 
cm$, $Z=-30 cm$.}
\label{fig.2}
\end{figure}

The corresponding quantum mean arrival time $\overline{\tau_Q}$ which is an 
observable quantity, is given by
\begin{eqnarray}
\overline{\tau_Q}=\frac{\int_{0}^{\infty}
{\Pi_{Q}}(t) t\, dt}{\int _{0}^{\infty}
{\Pi_{Q}}(t) dt}
\label{quantarrtim}
\end{eqnarray}
One may note that though the integral in the numerator of 
Eq.(\ref{quantarrtim}) formally 
diverges, several techniques have been employed in the literature ensuring 
rapid fall off for the probability distributions asymptotically\cite{hahne}, 
so that convergent results are obtained for the integrated arrival time. For 
our present purposes it is sufficient to employ the  simple strategy of taking
a cut-off ($t=tc$) in the upper limit of the 
time integral\cite{class,weq} with 
$tc=\sqrt{2(Z+6 \sigma_{tc} )/g}$
where $\sigma_{tc}=\sigma_T(t=tc)$ is the width of the wave packet at 
time $tc$. Thus,
our computations of the mean arrival times are valid up to the 
$6 \sigma$ level of
spread in the wave function. 

If we impose 
now the classical limit for the quantum TOF 
distribution 
given by Eq.(\ref{quantdistri}), then one can check that under the large mass 
and high temperature limits, and when $\sigma_0 << (kT/m)t^2 $, one can take 
${\sigma_T}^2 \approx (kT/m)t^2$. Thus, 
${\Pi_{Q}}(t)={\Pi_{C}}(t)$, i.e., the two distributions match 
in the limit of large mass and high temperature. The probability current
method of computing the quantum arrival time distribution furnishes an
effective way of approaching the classical limit of the distribution by
smoothly varying the parameters such as mass and temperature of the
quantum distribution\cite{class,weq}. Note also,
that the mass dependence of arrival time distribution given by 
Eq.(\ref{quantdistri}) and consequently, the observable mean arrival time
given by Eq.(\ref{quantarrtim}),  signifies the quantum mechanical
violation of the gravitational weak equivalence 
principle\cite{weq}. Thus TOF 
measurements\cite{expt,ion,dna,weiss,lett} offer a practical possibility for
experimental demonstration of the equivalence principle violation at the 
quantum level\cite{weq2}.

\section{Numerical Results}

We perform a numerical study
of the quantum arrival time distribution of the falling atomic cloud by
varying its mass and temperature separately.  
We first plot ${\Pi_{C}}(t)$ and ${\Pi_{Q}}(t)$ for fixed
temperature of $2.5\times 10^{-6}K$ in Fig.1. It is seen the classical and
the quantum distributions are clearly different for clouds of small atomic
mass such as Li and Na. However, as one increases the atomic mass, one sees
that 
${\Pi_{C}}(t)$ and ${\Pi_{Q}}(t)$ begin to overlap for heavy
atoms such as Rb for this value of temperature. The temperature variation
of the arrival time distributions are displayed in Fig. 2 where one sees
that even for heavy Rb atoms, the two distributions are quite distinct
in the low (nano kelvin) temperature range. It would be interesting if
our prediction in the low temperature and 
lower atomic mass region where the quantum 
time of flight distribution sharply differs from the classical TOF 
distribution could be verified in actual experiments.

\begin{figure}
\centering
\epsfxsize=3.0in\epsfysize=3.0in
\epsfbox{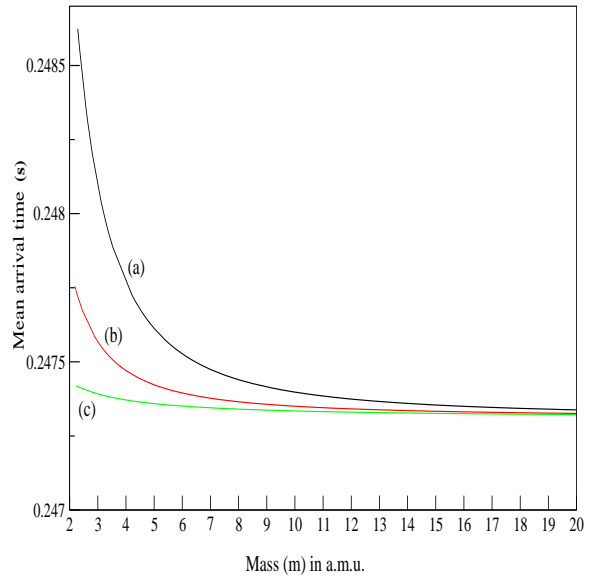}
\caption{\label{fig:epsart}
The mass variation of the mean arrival times $\overline{\tau_Q}$ and 
$\overline{\tau_C}$ are shown in the figure for a fixed value of 
temperature ($T=1.41 \times 10^{-6} K$) with $Z=-30~~cm$. The quantum 
mean arrival time $\overline{\tau_Q}$ is plotted for two different values
of $\sigma_0$. 2(a) $\overline{\tau_Q}$ for $\sigma_0=10^{-5} cm$, 2(b)
$\overline{\tau_Q}$ for $\sigma_0=2\times 10^{-5} cm$, 2(c) The mean 
arrival time $\overline{\tau_C}$ calculated through the classical 
TOF distribution ${\Pi_{C}}(t)$ with $T=1.41 \times 10^{-6} K$ and
$Z=-30~~cm$.}
\label{fig.3}
\end{figure}

The variation with mass of the quantum and classical 
mean arrival times at a 
particular detector location for the ensemble of falling atoms is depicted
in Fig.3. One can  see  that in the 
limit of large mass the mean arrival
time $\overline{\tau_Q}$ asymptotically approaches the classical result. 
One can also investigate 
the variation of the mean arrival times ($\overline{\tau_Q}$ 
and $\overline{\tau_C}$) with varying temperature of the cloud, and
obtain similar results, as expected from the temperature variation 
of the classical and quantum arrival time distributions plotted in Fig.2.   
The mass and temperature dependences
of the quantum arrival time distribution and consequently the quantum mean
arrival time arise essentially due to the spread of the wave packet for
the atoms. A smaller value of the initial width $\sigma_0$ for
the wave packet results in its faster spread. The amount of departure of the 
quantum distribution from 
its classical counterpart is thus contingent on the magnitude of the ensemble
spread (since ${\Pi_{C}}(t)$ is independent of $\sigma_0$). This is 
clearly depicted in Fig.4 where the classical (${\Pi_{C}}(t)$)
and quantum (${\Pi_{Q}}(t)$) TOF distributions are plotted for three 
different
values of $\sigma_0$ at a fixed mass and temperature. 

\begin{figure}
\centering
\epsfxsize=3.5in\epsfysize=4.5in
\rotatebox{0}{\epsfbox{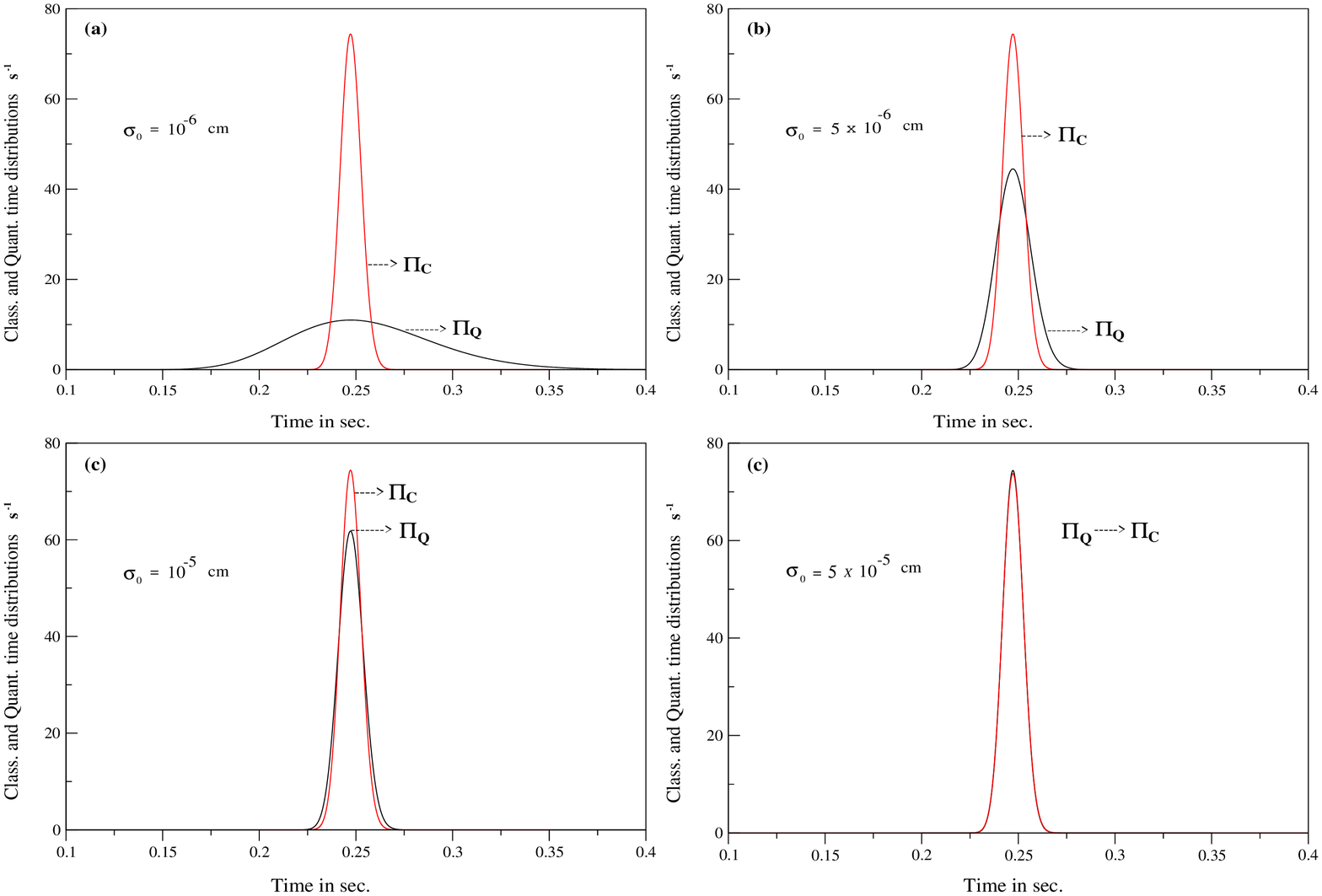}}
\caption{The classical (${\Pi_{C}}(t)$) and quantum (${\Pi_{Q}}(t)$)
TOF distributions of the atomic cloud falling freely under gravity are plotted
for four different values of $\sigma_0$ at a fixed mass of Be 
atom ($m=9.01 a.m.u$)
and at a fixed temperature $T=3.0 \times 10^{-6} K$  with $Z=-30 cm$.}
\label{fig.4}
\end{figure}

\section{Summary and Conclusions}

To summarize, in this work we have considered the analysis of the time of
flight measurements of falling cold atomic clouds. The inference of the 
temperature of the cloud in various experiments\cite{expt,ion,dna} is usually 
performed through a classical analysis in which the results obtained are
same as through the solution of Newton's equations for ballistic motion
of particles falling under gravity\cite{yavin}. Here we emphasize the relevance
of employing a quantum mechanical arrival time distribution for the analysis
of such experiments. We use the probability current density 
approach towards obtaining the arrival time or the 
TOF distribution.
Our definition of the quantum arrival time distribution and the
observable mean arrival time in terms of the 
modulus of the
probability current density is particularly motivated from the equation
of continuity, and other physical considerations discussed in the
literature\cite{current,bohmarrtim,holland,ali,cpviol}.  Further, we generalise
the probability current density approach to calculate the
quantum arrival time distribution for a mixed state describing the 
Maxwell-Boltzmann distribution of velocities for the falling atomic clouds
in the relevant experiments\cite{expt,weiss,lett}. We compute the TOF
distribution and mean arrival time through this scheme and compare our 
results with those obtained through a classical analysis\cite{tof} for
various atomic masses. 

The obtained quantum 
arrival time distribution matches with the classical TOF distribution in
the high temperature and the large mass limits, hence furnishing another 
example
of the smooth emergence of the classical limit\cite{class}
of the quantum arrival time in the framework of the
probability current approach. However, a clear distinction between the 
quantum and the classical distributions is exhibited for either small atomic
mass, or low temperature of the cloud. This results from differential 
wave packet spreading depending upon the mass,
velocity and width of the wave packet for the atoms.  
Our scheme thus provides a new method for experimental verification of the
probability current density approach for calculating the arrival time
distribution, in addition to an earlier proposed method using the spin
rotator as a quantum clock\cite{spinrot}.
Finally, we wish to emphasize that more investigations
of modern experiments employing time of flight techniques should be performed
using various quantum mechanical schemes\cite{1}. Such studies have the
potential to empirically resolve ambiguities inherent in the theoretical
formulations of the quantum arrival time distribution using  
cold trapped atom experimental techniques, and may also shed new light on the 
inference of the experimental data.

{\bf Acknowledgments}               

This work was initiated while MMA was visiting the Atominstitut der 
${\ddot O}$sterreichischen Universit${\ddot a}$ten, Vienna. MMA 
would like to thank Yuji Hasegawa and  Helmut Rauch for the stimulating 
discussions 
and warm hospitality at the Atomic Institute. DH acknowledges the support 
of the Jawaharlal Nehru Fellowship.


\begin{thebibliography}{99}
\bibitem{expt} See, for example, M. Schellekens et al., Science {\bf 310}, 648 
(2005); A. Ottl et al. Phys. Rev. Lett. {\bf 95}, 090404 (2005).
\bibitem{ion} E. Moskovets and A. Vertes, Rapid Commun. Mass Spectrom. 
{\bf 13}, 2244 (1999).
\bibitem{dna}  J. M. Butler, P. Jiang-Baucom, M. Huang, P. Belgrader and
J. Girard, Anal. Chem. {\bf 68}, 3283 (1996).
\bibitem{weiss} D. S. Weiss, E. Riis, Y. Shevy, P.J. Ungar, and S. Chu,
J. Opt. Soc. Am. B {\bf 6}, 2072 (1989).
\bibitem{yavin} I. Yavin, M. Weel, A. Andreyuk, and A. Kumarakrishnan,
Am. J. Phys. {\bf 70}, 149 (2002).
\bibitem{tof} D. Bassi, U. Hefter, K. Bergmann, D.J. Auerbach and M. Zen  in: 
{\it Atomic and Molecular Beam Methods}, G.Scoles (Ed.) (Oxford University 
Press, Oxford, 1988); C. Grupen, {\it Particle Detectors}, (Cambridge 
University Press, 
Cambridge, 1996); Ch. Kurtsiefer and J.Mlynek, Appl. Phys. 
B {\bf 64}, 85 (1997); Ch. Kurtsiefer, T. Pfau and J. Mlynek, Nature {\bf 386},
150 (1997). 
\bibitem{kijowski}
J. Kijowski, Rept. Math. Phys. {\bf 6}, 351 (1974).
\bibitem{operator}
N. Grot, C. Rovelli and R. S. Tate, Phys. Rev. A{\bf 54}, 4676 (1996); V. 
Delgado and J. G. Muga, Phys. Rev. A{\bf 56}, 3425 (1997).
\bibitem{bohmarrtim}
W. R. McKinnon  and C. R. Leavens, 
Phys. Rev. A {\bf 51}, 2748 (1995); C. R. Leavens, Phys. Rev. A {\bf 58}, 840 
(1998).
\bibitem{1} J. G. Muga and C. R. Leavens, Phys. Rep\emph{.} {\bf 338}, 353
(2000); 
\emph{Time in Quantum Mechanics}, edited by J. G. Muga, R. 
Sala Mayato
and I. L. Egusquiza (Springer-Verlag, Berlin, 2002).
\bibitem{current} R. S. Dumont  and T. L. Marchioro II,  Phys. Rev. A 
{\bf 47}, 85 (1993); C. R. Leavens,  Phys. Lett. A {\bf 178}, 27 (1993); 
J. G. Muga, S. Brouard   and D. Macias,  Ann. Phys. {\bf 240}, 351 (1995);  
A. Challinor, A. Lasenby, A. S. Somaroo, C. Doran and S. Gull,  Phys. Lett. A 
{\bf 227}, 143 (1997);  V. Delgado,  Phys. Rev A {\bf 59}, 1010 (1999).
\bibitem{holland}
P. Holland, Phys. Rev. A {\bf 60}, 4326 (1999); 
P. Holland, Ann. Phys. (Leipzig) {\bf 12}, 446 (2003); 
P. Holland and C. Philippidis Phys. Rev. A {\bf 67}, 062105 (2003).
\bibitem{ali}
Md. M. Ali, A. S. Majumdar, D. Home and S. Sengupta Phys. Rev. A {\bf 68},
042105 (2003).
\bibitem{class} Md. M. Ali, A. S. Majumdar and A. K. Pan, Found. Phys. Lett. 
{\bf 19}, 723 (2006).  
\bibitem{weq} Md. M. Ali, A. S. Majumdar, D. Home and A.K. Pan, Class. 
Quantum Grav. {\bf 23}, 6493 (2006).
\bibitem{lett} P.D. Lett, W.D. Phillips, S.L. Rolston, C.E. Tanner, R.N. Watts,
and C.I. Westbrook, J. Opt. Soc. Am. B {\bf 6}, 2084 (1989).
\bibitem{finkelstein}
J. Finkelstein, Phys. Rev. A{\bf 59}, 3218 (1998).
\bibitem{baere}
W. Struyve, W. D. Baere, J. D. Neve and S. D. Weirdt, Phys. Lett. A
{\bf 322}, 84 (2004).
\bibitem{hahne}
J. A. Damborenea, I. L. Egusquiza and J. G. Muga, eprint quant-ph/0109151;
G. E. Hahne,  J. Phys. A {\bf 36}, 7149 (2003).
\bibitem{weq2} D. M. Greenberger, Rev. Mod. Phys. {\bf 55}, 875 (1983).
\bibitem{cpviol}
D. Home and A. S. Majumdar,
Found. Phys. {\bf 29}, 721 (1999); A. S. Majumdar and D. Home,
Phys. Lett. A {\bf 296}, 176 (2002).
\bibitem{spinrot}
A. K. Pan, Md. M. Ali and D. Home, Phys. Lett. A{\bf 352}, 296 (2006).

\end{thebibliography}
\end{document}